\documentclass{emulateapj}

\usepackage{psfig}
\usepackage{amsmath}
\usepackage{amssymb}
\usepackage{graphicx}

\newcommand{\kms}{\ensuremath{\mathrm{km~s}^{-1}}}

\newcommand{\msun}{\ensuremath{M_\odot}}
\newcommand{\Matm}{\ensuremath{m_{\mathrm{atm}}}}

\shortauthors{Kasen \& Plewa}
\shorttitle{Spectral Signatures of GCD}

\begin{document}

\title{Spectral Signatures of Gravitationally Confined Thermonuclear Supernova 
Explosions}

\author{Daniel Kasen\altaffilmark{1,2,3} and Tomasz Plewa\altaffilmark{4,5,6}}
\altaffiltext{1}{Allan C. Davis Fellow, Department of Physics and Astronomy, Johns Hopkins University,
Baltimore, MD 21218}
\altaffiltext{2}{Space Telescope Science Institute, Baltimore, MD 21218}
\altaffiltext{3}{Lawrence Berkeley National Laboratory, Berkeley, CA 94720}
\altaffiltext{4}{Center for Astrophysical Thermonuclear Flashes,
   The University of Chicago,
   Chicago, IL 60637}
\altaffiltext{5}{Department of Astronomy \& Astrophysics,
   The University of Chicago,
   Chicago, IL 60637}
\altaffiltext{6}{Nicolaus Copernicus Astronomical Center,
   Bartycka 18,
   00716 Warsaw, Poland}

\begin{abstract} 

We consider some of the spectral and polarimetric signatures of the
gravitational confined detonation scenario for Type~Ia supernova
explosions.  In this model, material produced by an off-center
deflagration (which itself fails to produce the explosion) forms a
metal-rich atmosphere above the white dwarf surface.  Using
hydrodynamical simulations, we show that this atmosphere is compressed
and accelerated during the subsequent interaction with the supernova
ejecta.  This leads ultimately to the formation of a high-velocity
pancake of metal-rich material that is geometrically detached from the
bulk of the ejecta.  When observed at the epochs near maximum light,
this absorbing pancake produces a highly blueshifted and polarized
calcium IR triplet absorption feature similar to that observed in
several Type~Ia supernovae.  We discuss the orientation effects
present in our model and contrast them to those expected in other
supernova explosion models.  We propose that a large sample of
spectropolarimetric observations can be used to critically evaluate
the different theoretical scenarios.

\end{abstract}


\section{Introduction}
After decades of study, the mechanism of the explosion of a white
dwarf in a Type~Ia supernovae remains uncertain
\citep{Hillebrandt_Niemeyer}.  Several of the proposed theoretical
models \citep{Arnett_Detonation,Khokhlov_def, Reinecke_3D} have failed
to produce objects with the energetics and chemical composition
compatible with observations \citep{Hoeflich_99by}. These two
characteristics are best captured by the so-called delayed-detonation
models \citep{Khokhlov_DD, Gamezo_DDT, Plewa_GCD}.  In such models, a
mild ignition occurs near the center of the accreting, massive white
dwarf, and sparks a deflagration (subsonic flame). During the
subsequent evolution, a deflagration to detonation (supersonic
reactive wave) transition (DDT) takes place.

Despite the promise of the DDT models, it remains unclear how a
transition to detonation occurs -- in all standard DDT models
calculated so far, the detonation was triggered artificially.
However, \cite{Plewa_GCD} have proposed a gravitational confined
detonation (GCD) model in which a detonation naturally follows a
slightly off-center ignition.  In this scenario, the deflagration
takes the form of a single bubble, buoyantly rising to the stellar
surface.  At the bubble's breakout, the surface layers of the star are
laterally accelerated and begin sweeping across the stellar surface,
converging opposite the breakout point.  The subsequent compression of
the colliding streams and thermalization of the kinetic energy triggers
a detonation.  The authors speculate that, as in the standard DDT
case, the detonation will consume the remaining fuel, producing an
energetic explosion with chemically stratified ejecta. The stability
of the GCD is currently the subject of more detailed numerical study.

One of the most striking properties of the GCD model is that the
products of the deflagration are brought to the surface of the white
dwarf prior to detonation.  This material constitutes a small fraction
of the stellar mass, and is rich in metals.  This compositional
``pollution'' of the outer stellar layers is reminiscent of a peculiar
feature noticed in the spectra of some Type~Ia~SNe.  In a handful of
objects, observers have identified a highly-blueshifted Ca~II IR
triplet absorption \citep{Hatano_94D, Li_00cx, Wang_01el,
Gerardy_03du}, indicating absorbing material moving at velocities
$\sim 20,000$~\kms, much higher than that characteristic of other
spectral lines.  Given that velocity is proportional to radius in
expanding supernova atmospheres, the high-velocity (HV) calcium
absorption indicates a component of absorbing material geometrically
detached from the bulk of the ejecta.  Spectropolarimetry of SN~2001el
further showed that the HV absorption was highly polarized, indicating
that the absorbing material was distributed aspherically with perhaps
a clump-like geometry \citep{Wang_01el, Kasen_01el}.

In this letter we demonstrate that the ejecta structure characteristic
of the GCD scenario can naturally explain observations of the peculiar
HV calcium absorption feature in Type~Ia supernovae.

\section{Methods}
\label{Methods_Sec}

\begin{figure*}[ht]
\begin{center}
\includegraphics[height=5.5cm,clip=true]{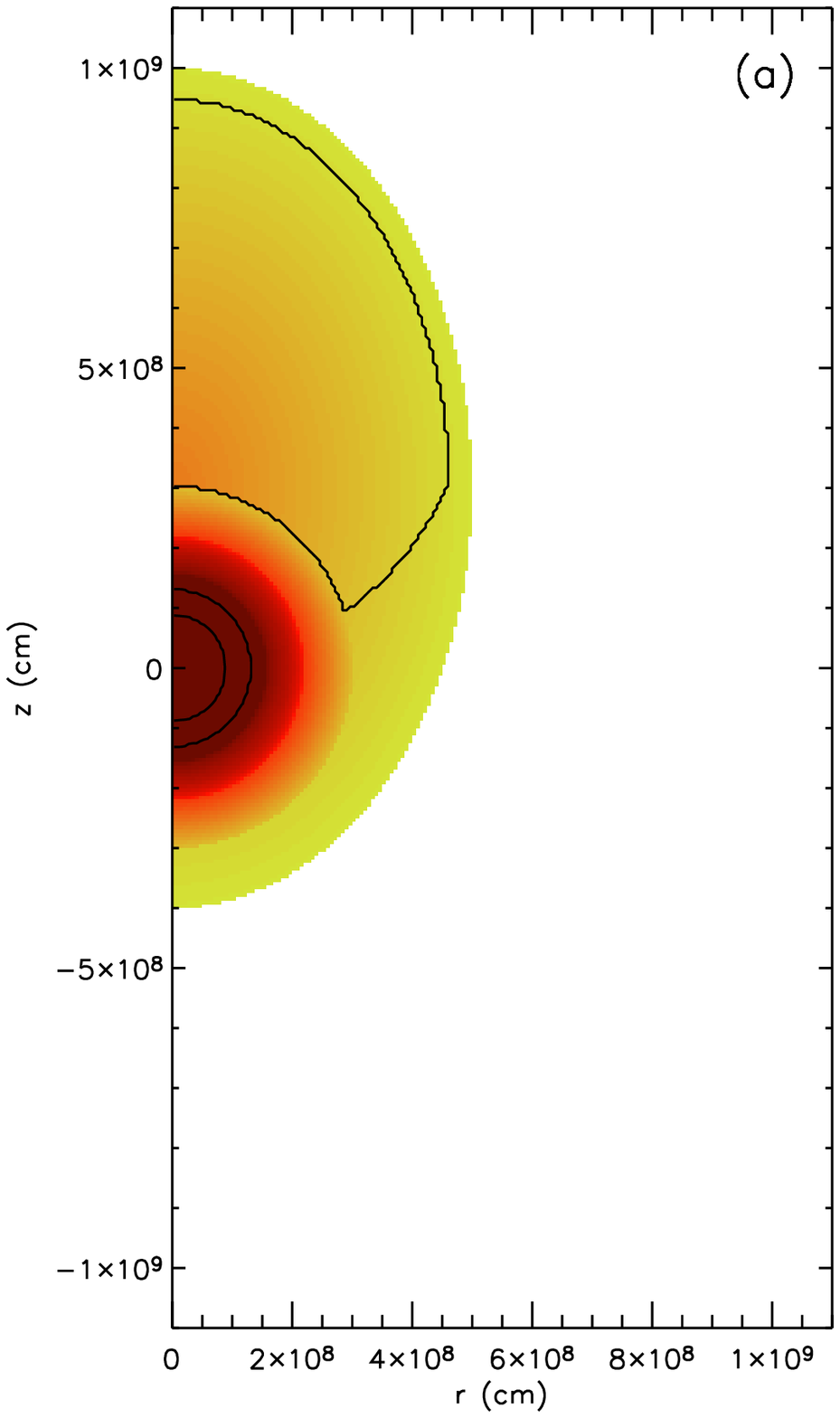}%
\includegraphics[height=5.5cm,clip=true]{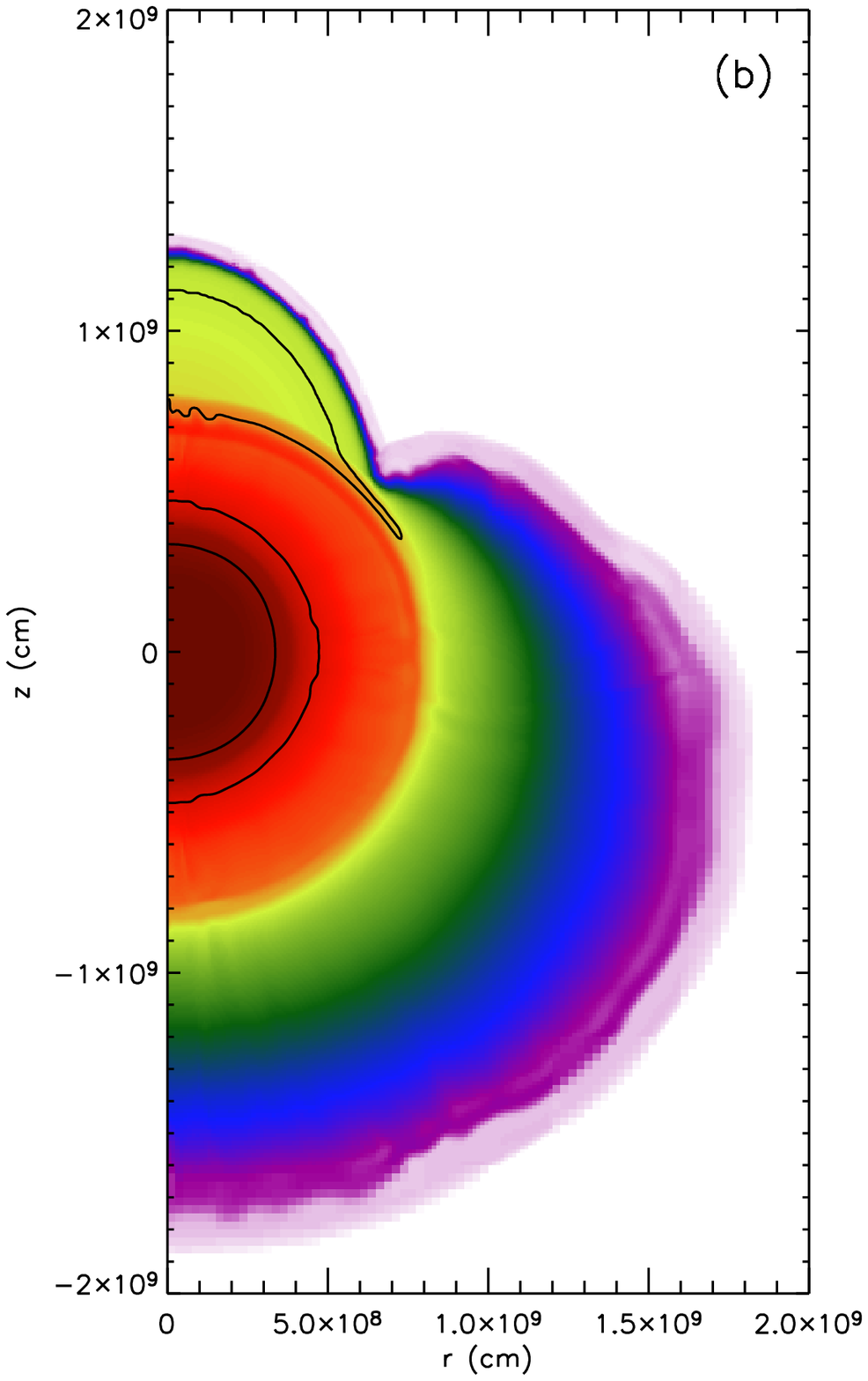}%
\includegraphics[height=5.5cm,clip=true]{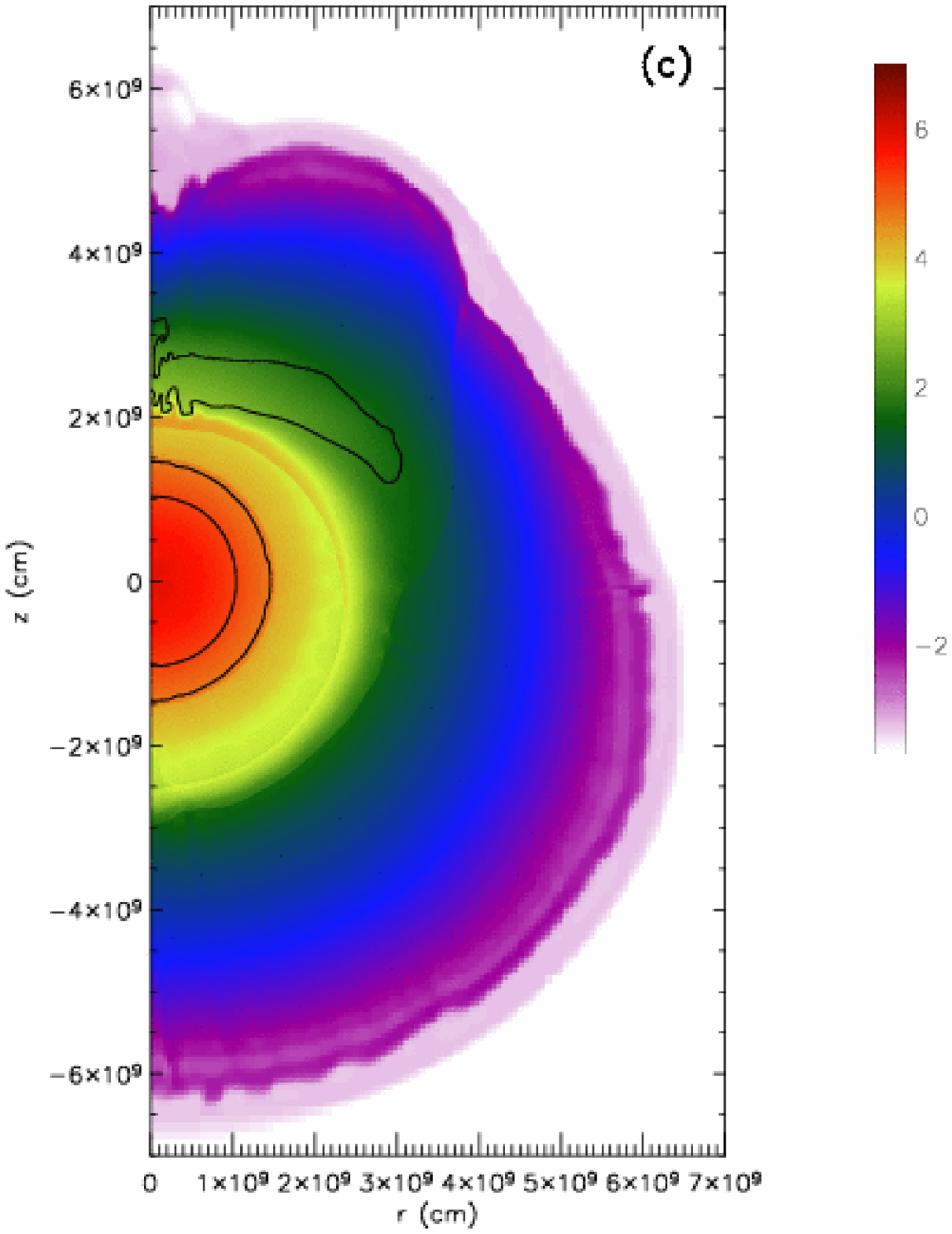}%
\includegraphics[height=5.5cm,clip=true]{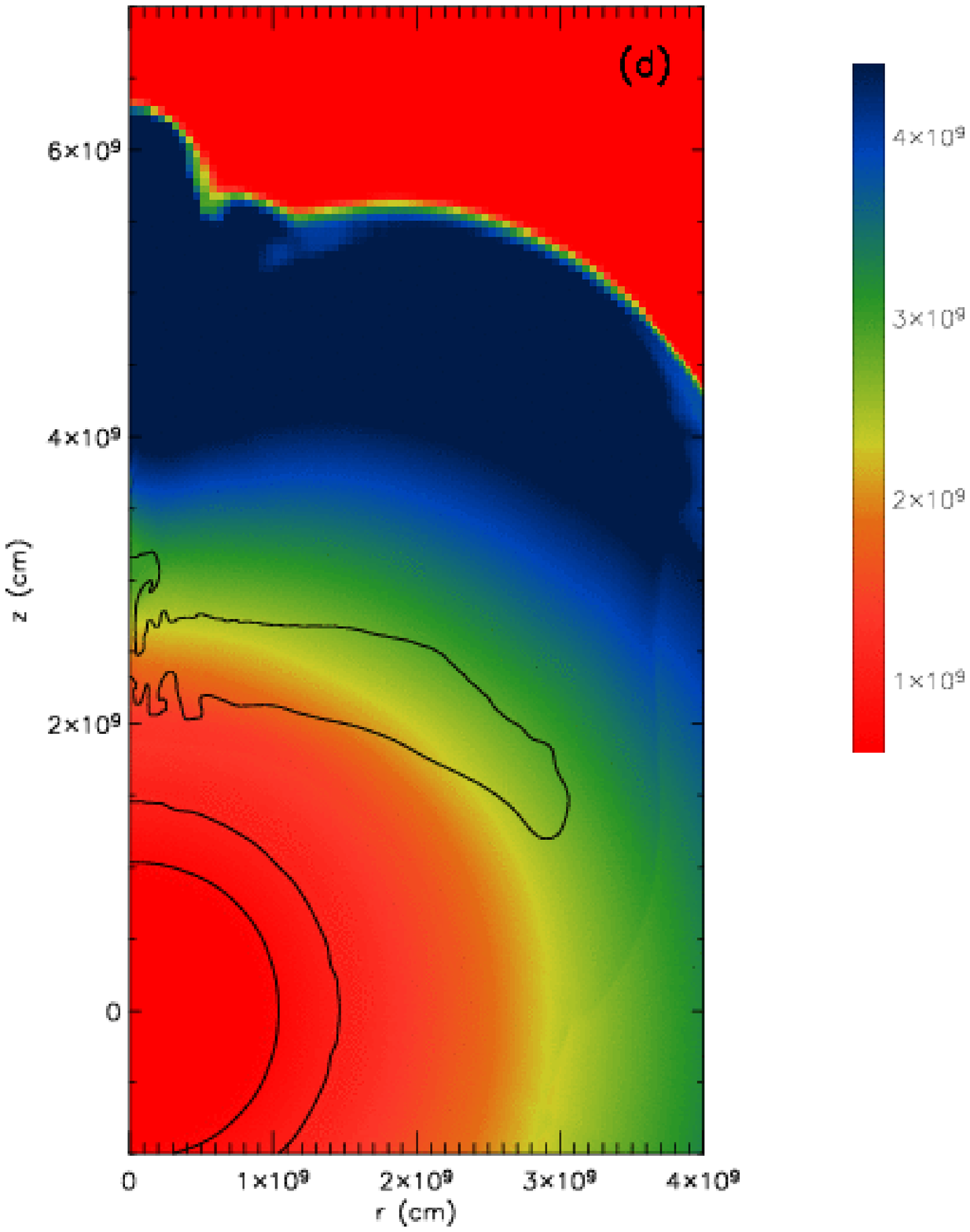}
\end{center}
\caption{Hydrodynamical simulation of the model supernova ejecta
interacting with the extended atmosphere.  Panels~(a)-(c) show the
density in log scale; panel~(d) depicts the magnitude of the
velocity.  The density scale is shown by the colorbar in panel~(c)
while the velocity scale is shown by the colorbar in panel~(d). The
black contour line corresponds to a calcium number abundance of 0.01.
(a) Density distribution at the beginning of the simulation
($t=0$~s).  (b) Density distribution at $t=0.3$~s.  Notice the
remarkable deformation of the calcium-rich material in the upper part
of the computational domain. (c) Density distribution at the
final time ($t=1.24$~s).  Notice that the calcium-rich region has been
strongly compressed into a pancake-like structure.  (d): Velocity
magnitude at the final time.  The calcium-rich absorber is seen moving
at velocity $\sim 21,000$~\kms\ with a substantial velocity
gradient across the structure.}
\label{hydro_fig}
\end{figure*}

We studied the post-detonation evolution of the GCD model using a
simplified numerical setup in which spherical supernova ejecta runs
into an aspherical metal-rich atmosphere.  For the supernova ejecta
structure, we used the W7 model \citep{Nomoto_w7} with an initial
radius of $3\times10^8$~cm.  Around the ejecta, we placed an
ellipsoidal metal-rich extended atmosphere representing the bubble of burned
material expelled during the GCD breakout.  The composition of this
material was taken to be the oxygen-burned composition \#4 from
Table~3 of \cite{Khokhlov_LC} which consists of 57\% silicon, 27\%
sulfur, 7.1\% iron, 2.7\% calcium and small amounts of other elements.
The extended atmosphere had an axis ratio of 1.2, semi-major axis length 
of $6\times 10^8$~cm  and was centered at $3\times 10^8$~cm.  The density of
the atmosphere decreased exponentially with a scale length of
$3\times10^8$~cm.  The atmosphere was uniformly expanding about its
center with velocity proportional to radius and reaching 5000~\kms\ at
the outer edge.  The ejecta and ellipsoidal extended atmosphere were
embedded in an ambient medium of pure helium of density
$10^{-4}$~g~cm$^{-3}$.  The initial model was isothermal with $T =
10^7$~K.  This overall configuration closely represents that seen in
the GCD calculations of \cite{Plewa_GCD}.

The hydrodynamic evolution was calculated with the adaptive mesh
refinement code {\sc flash} \citep{Fryxell_flash}.  The initial model was
defined on a 2-D cylindrical grid covering the region up to
$2.56\times10^{10}$~cm in radius and from $-2.56\times 10^{10}$~cm to
$2.56\times 10^{10}$~cm in the z-direction.  The ejecta was centered
at the origin of the grid.  The maximum resolution of the simulation,
used only to resolve strong flow structures in the dense regions of
the model, was equal to 62.5~km.  We used a reflecting boundary
condition at the symmetry axis and allowed for free outflow otherwise.
We used a Helmholtz equation of state, an iso13 composition
\citep{Fryxell_flash}, and a multipole solver with 10 terms in the
expansion to account for self-gravity.

Spectra of the hydrodynamical models were calculated using a
multi-dimensional Monte Carlo radiative transfer code
\citep{Thomas_Thesis,Kasen_PhD}.  The opacities used in the
calculation were electron scattering and bound-bound line transitions.
Ionization and excitation were computed assuming local thermodynamic
equilibrium, where the temperature structure of the atmosphere was
determined self-consistently using an iterative approach enforcing
radiative-equilibrium \citep{Lucy_Radeq}.  Line interactions were
treated in the Sobolev approximation and included both absorption and
scattering according to a equivalent two level atom scheme with
thermalization parameter $\epsilon= 0.01$.  Monte Carlo photon packets
were initially emitted from a spherical inner boundary surface located
at $v = 6000$~\kms\ and according to a blackbody distribution with $T
= 9000$~K.  The packets were initially unpolarized but acquired
polarization by electron scattering. Line scattered light was assumed
to be unpolarized due to complete redistribution.  To construct the
emergent spectra and polarization, escaping photon packets were
collected into one of 100 angular bins.

\section{Results}
\label{Results_Sec}

We have calculated several models assuming different mass of the
extended atmosphere.  Here we discuss the evolution for one case in
detail ($\Matm \simeq 0.008$~\msun) and describe the differences in
the other cases as appropriate.  

The initial setup of our model is shown in Fig.~\ref{hydro_fig}(a).
The W7 ejecta occupies the innermost region of the grid, shown in
shades of red.  The yellow ellipsoidal region surrounding the ejecta
represents material expelled during the deflagration.  The black
contour lines enclose the regions where the calcium number abundance
exceeds 1\%. In particular, the contour in the top portion of the
image marks calcium produced in the deflagration; the inner
calcium-rich ring is that of the W7 nucleosynthesis.

The overall hydrodynamic evolution in our model is best described as
ejecta impacting a stratified atmosphere, with shock-related effects
playing a minor role. In Fig.~\ref{hydro_fig}(b) ($t=0.3$~s), a mostly
spherical expansion of the supernova shock can be seen in the lower
part of the computational domain.  In the upper part of the domain,
the ejecta impacts and deforms the extended atmosphere.  By $t=1.24$~s
(Fig.\ref{hydro_fig}(c)), that impact has lead to the compression of
the calcium-rich material into a pancake-like structure of radius
$\sim 30,000$~km, located near $z \approx 2\times10^9$~cm.  By that
time, the ambient medium surrounding the atmosphere has been
completely overrun by the forward supernova shock.  The reverse shock
can be seen as the highly contrasted spherical structure starting at
$(r,z) = (0,-2.6\times10^9)$~cm.

Beginning at $t\approx1$, the overall expansion becomes increasingly
homologous, and we stop our calculation at $t=1.24$~s.
Fig.~\ref{hydro_fig}(d) shows the total velocity in the upper portion
of the computational domain at the final time.  The calcium-rich
pancake structure shows a significant velocity gradient across its
body ($\Delta v/c \approx 0.02$).  The central region of the pancake
$(r=0$) moves with velocity spanning 17,000-24,000~\kms, while near
the outer rim the velocity is higher (22,000-28,000~\kms).

In the case of a more massive atmospheres, the overall evolution and
resulting morphology are similar to the case described above. The
major difference is the final velocity of the calcium-rich pancake.
For a model with $\Matm = 0.016$~\msun, the inner part of the
pancake moves with velocities 14,000-20,000~\kms, while the outer edge
moves at 21,000-26,000~\kms.  For an even more massive atmosphere
($\Matm = 0.08$~\msun) the velocities are still lower, increasing
from 10,000-12,000~\kms to 17,000-21,000~\kms across the pancake.  In
this case the calcium-rich pancake almost overlays the region of
intermediate mass elements in the W7 ejecta.

\begin{figure}[t]
\begin{center}
\psfig{file=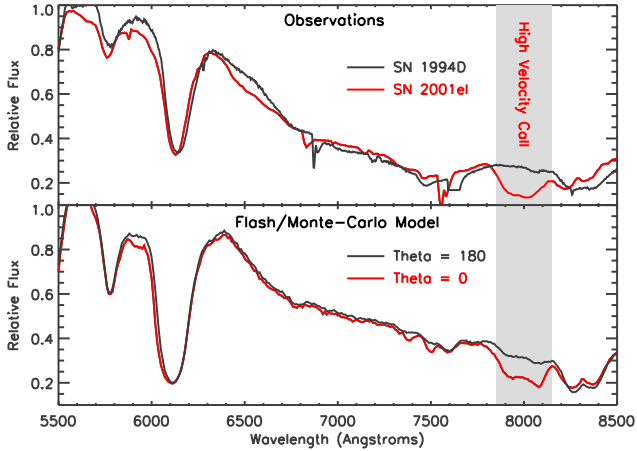,height=2.4in}
\caption{Comparison of the synthetic model spectra to Type~Ia
supernova observations.  Top panel: Spectral observations of two
Type~Ia SNe near maximum light; SN~2001el \citep[red
line,][]{Wang_01el} shows a strong HV calcium absorption at 8000~\AA\
while SN~1994D \citep[black line,][]{Patat_94D} does not.  Bottom
panel: Synthetic model spectra at 20 days.  A HV calcium absorption is
clearly seen when looking straight down on the calcium-rich pancake
(viewing angle $\theta = 0^\circ$, red line), whereas none is seen from the
opposite side ($\theta = 180^\circ$, black line).
\label{spec_fig}}
\end{center}
\end{figure}

\begin{figure}[t]
\begin{center}
\psfig{file=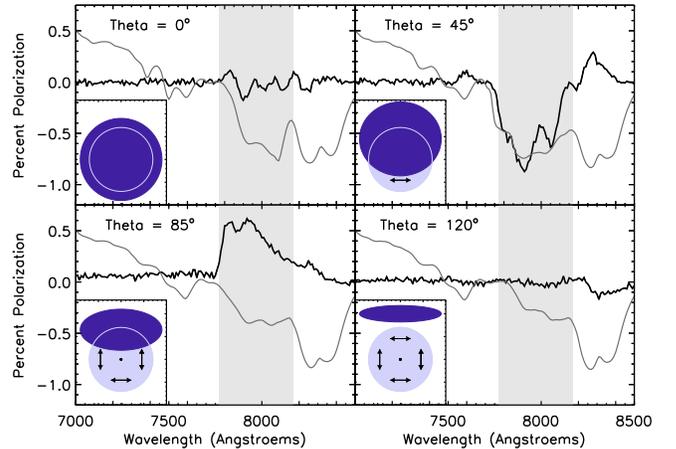,height=2.4in}
\caption{Synthetic polarization spectra of the model at 20 days.  By
convention, a positive (negative) polarization level signifies
polarization aligned parallel (perpendicular) to the symmetry axis.
The black lines show the polarization level while the gray lines show
the corresponding (arbitrarily scaled) flux spectrum.  The insets
illustrate how the HV calcium line polarization is caused by the
partial obscuration of the supernova photosphere (light blue disk)
by the calcium-rich pancake (dark blue object).
\label{pol_fig}}
\end{center}
\end{figure}

The synthetic spectrum of the model at a time when the supernova is
near maximum light has been obtained after homologously expanding
the ejecta to 20~days (lower panel of Fig.~\ref{spec_fig}).  Given the
enhanced abundance of intermediate mass elements and relatively low
temperature ($T \approx 5500$~K), the pancake is opaque in the Ca~II
IR triplet lines.  For a viewing angle, $\theta$, in which the
observer looks directly down upon the pancake ($\theta = 0^\circ$),
the pancake obscures the supernova photosphere, creating a broad and
highly blueshifted HV calcium absorption feature near 8000~\AA.  The
HV feature seen in the model compares well to that observed in
SN~2001el \citep{Wang_01el}, shown in the top panel of
Fig.~\ref{spec_fig}. For larger viewing angles, the pancake obscures
less (or none) of the photosphere, and the HV feature is weaker or
absent in the model spectrum.  This orientation effect may explain why
in some supernovae, such as SN~1994D \citep{Patat_94D}, a HV calcium
feature at maximum light is seen only weakly or not at all.

The aspherical geometry of the calcium-rich pancake also leads to
significant polarization over the HV calcium feature
(Fig.~\ref{pol_fig}).  Light in the supernova ejecta becomes polarized
by electron scattering.  By convention, a positive (negative)
polarization level signifies polarization oriented parallel
(perpendicular) to the symmetry axis. Because the electron-scattering
photosphere in our model is essentially spherical, the observed
polarization cancels in the continuum.  However, the HV pancake may
partially obscure the underlying photosphere, leading to a non-zero
polarization over the HV calcium feature.  This effect is illustrated
by the insets to Fig.~\ref{pol_fig}.

The line polarization created in this way depends sensitively upon the
viewing angle.  For $\theta = 0^\circ$, the polarization cancels
completely due to the azimuthal symmetry (Fig.~\ref{pol_fig}(a)).  As
$\theta$ is increased, the absorber selectively reveals horizontally polarized
light from the lower rim of the photosphere, resulting in negative
line polarization of order $\sim 1\%$ (Fig.~\ref{pol_fig}(b)).  For
$\theta = 45^\circ$, the model closely reproduces the HV line
polarization peak observed in SN~2001el \citep{Wang_01el}.  For larger
$\theta$, vertically polarized light from the side edges of the
photosphere dominates, and the line polarization changes sign
(Fig.~\ref{pol_fig}(c)).  For $\theta \ga 110^\circ$, the pancake no
longer obscures the photosphere, and leaves no obvious signature in
either the flux or polarization spectrum (Fig.~\ref{pol_fig}(d)).

\section{Discussion}
\label{Discussion_Sec}

We have demonstrated that the observations of a peculiar HV calcium
feature in Type~Ia supernovae are naturally explained within the GCD
model.  In the model, burned material expelled during the breakout of
the deflagrating bubble forms an extended atmosphere above the stellar
surface.  That atmosphere is subsequently compressed and accelerated
during the hydrodynamic interaction with the supernova ejecta.  This
leads to the formation of a high-velocity calcium-rich pancake with
cross-section comparable to that of the underlying photosphere. The
partial obscuration of the photosphere by the pancake results in a HV
calcium absorption feature with significant line polarization.

In our model, the blueshift of the absorption feature depends
sensitively upon the mass of the expelled material.  For the fiducial
case studied here ($\Matm = 0.008~\msun$), the pancake material spans
the velocity range 17,000-24,000~\kms, and is geometrically detached
from the bulk of the supernova ejecta.  This compares well with the
velocities inferred from the HV calcium feature of SN~2001el.  As the
atmosphere mass is increased, the absorbing pancake moves at lower
velocity and eventually blends with the region of intermediate mass
elements in the supernova ejecta.  In such a case we might expect a
very different observable signature, in which the pancake material
increases the strength and blueshift of several of the normal Type~Ia
spectral features.  This could provide an orientation-dependent
explanation for the unusually high velocities and peculiar velocity
evolution of the normal features in some Type~Ia SNe (e.g., SN~1984A
\citep{Branch_84A} and SN~2002bo \citep{Benetti_02bo}).

The large size of the absorbing pancake results in the distinctive
orientation effects present in our model.  The strength of the HV flux
absorption depends only upon the fraction of the photosphere obscured
by the pancake, and hence decreases monotonically with $\theta$. The
line polarization, in contrast, is maximal when either a large or a
small part of the photosphere is covered.  Such behavior could be
tested with a large sample of spectropolarimetric observations of
SNe~Ia; in particular, one could study the correlation between the HV
calcium polarization level with the depth of the flux absorption.

Although the HV calcium feature is the most profound signature of the
absorbing pancake, the opacity due to numerous iron and titanium lines
creates a modest flux depression in the wavelength region
3500-4500~\AA.  This leads to a variation of the B-band magnitude with
viewing angle of order one tenth of a magnitude.  As discussed in
\cite{Branch_00cx} and \cite{Thomas_00cx}, the temporal evolution of
this opacity should also affect the shape of the supernova light
curve.  Therefore, the presence of the calcium-rich absorber is
another intrinsic source of SNe~Ia photometric diversity, of possible
relevance to their cosmological application.

The presence of a HV calcium absorber can possibly be explained in
other explosion scenarios.  For example, the large bubbles of burned
material present in the pure deflagration and standard DDT models may
in principle produce a number of HV absorbers.  This case can be
distinguished from the case of single absorber in the GCD model using
a rich sample of spectropolarimetric observations. In particular, our
model predicts a much smaller fraction of supernovae showing a
persistent HV calcium feature, as well as the aforementioned
polarization correlations.  Note that multi-epoch observations may be
necessary to separate the signatures of the HV absorber from the
``transient ionization'' effect discussed by \cite{Gerardy_03du},
whereby the recombination of Ca~III to Ca~II in the cold outer layers
of ejecta may also give rise to a (short-lived) HV IR triplet
absorption in the epochs prior to maximum light.

For several SNe~Ia, polarization in the continuum has also been
detected, indicating a global asymmetry in the bulk of the supernova
ejecta.  In SN~2001el, the continuum polarization angle differed from
that of the HV calcium feature, suggesting that the orientation of the
HV absorber deviated from that of the ejecta.  One possible
explanation for this difference is that the ejecta acquired a
separate, large-scale asymmetry due to the interaction with a
companion star \citep{Marietta,Kasen_hole}.  Another interesting
possibility is to consider the GCD framework with rotation of the
progenitor included.  In such a case, the trajectory of the buoyantly
rising deflagrating bubble is not expected to be aligned with the
rotation axis of the progenitor.  This lack of correlation follows
from the fact that the convective core of the white dwarf is expected
to produce ignition seed points in a largely stochastic manner.  Such
a scenario should be studied in the future by means of integrated
multi-dimensional hydrodynamical simulations.

\acknowledgements
We thank Lifan Wang for use of the spectra of
SN~2001el, Rollin Thomas for collaborating in the development of the
radiative transfer techniques, and the anonymous referee for helpful
comments. This work is supported in part by the U.S. Department of
Energy under Grant No.\ B523820 to the Center for Astrophysical
Thermonuclear Flashes at the University of Chicago.  DK acknowledges
support from a NASA ATP grant.  This research used resources of the
National Energy Research Scientific Computing Center, which is
supported by the Office of Science of the U.S. Department of Energy
under Contract No.  DE-AC03-76SF00098.


\begin{thebibliography}{26}
\expandafter\ifx\csname natexlab\endcsname\relax\def\natexlab#1{#1}\fi

\bibitem[{{Arnett}(1969)}]{Arnett_Detonation}
{Arnett}, D.~W. 1969, Ap\&SS, 5, 180

\bibitem[{{Benetti} {et~al.}(2004)}]{Benetti_02bo}
{Benetti}, S. {et~al.} 2004, \mnras, 348, 261

\bibitem[{{Branch}(1987)}]{Branch_84A}
{Branch}, D. 1987, ApJ, 316, L81

\bibitem[{{Branch} {et~al.}(2004)}]{Branch_00cx}
{Branch}, D. {et~al.} 2004, ApJ, 606, 413

\bibitem[{{Fryxell} {et~al.}(2000)}]{Fryxell_flash}
{Fryxell}, B. {et~al.} 2000, ApJS, 131, 273

\bibitem[{{Gamezo} {et~al.}(2004){Gamezo}, {Khokhlov}, \& {Oran}}]{Gamezo_DDT}
{Gamezo}, V.~N., {Khokhlov}, A.~M., \& {Oran}, E.~S. 2004, PRL, 92, 211102

\bibitem[{{Gerardy} {et~al.}(2004)}]{Gerardy_03du}
{Gerardy}, C.~L. {et~al.} 2004, ApJ, 607, 391

\bibitem[{{H{\" o}flich} {et~al.}(2002)}]{Hoeflich_99by}
{H{\" o}flich}, P. {et~al.} 2002, ApJ, 568, 791

\bibitem[{{Hatano} {et~al.}(1999)}]{Hatano_94D}
{Hatano}, K. {et~al.} 1999, ApJ, 525, 881

\bibitem[{{Hillebrandt} \& {Niemeyer}(2000)}]{Hillebrandt_Niemeyer}
{Hillebrandt}, W. \& {Niemeyer}, J.~C. 2000, ARA\&A, 38, 191

\bibitem[{Kasen(2004)}]{Kasen_PhD}
Kasen, D. 2004, PhD thesis, University of California, Berkeley

\bibitem[{{Kasen} {et~al.}(2003)}]{Kasen_01el}
{Kasen}, D. {et~al.} 2003, ApJ, 593, 788

\bibitem[{{Kasen} {et~al.}(2004)}]{Kasen_hole}
---. 2004, ApJ, 610, 876

\bibitem[{Khokhlov(1991)}]{Khokhlov_DD}
Khokhlov, A. 1991, A\&A, 245, 114

\bibitem[{{Khokhlov} {et~al.}(1993){Khokhlov}, {M\"uller}, \&
  {H\"oflich}}]{Khokhlov_LC}
{Khokhlov}, A., {M\"uller}, E., \& {H\"oflich}, P. 1993, A\&A, 270, 223

\bibitem[{{Khokhlov}(2001)}]{Khokhlov_def}
{Khokhlov}, A.~M. 2001, ApJ, submitted; astro-ph/0008463

\bibitem[{{Li} {et~al.}(2001)}]{Li_00cx}
{Li}, W. {et~al.} 2001, PASP, 113, 1178

\bibitem[{{Lucy}(1999)}]{Lucy_Radeq}
{Lucy}, L.~B. 1999, A\&A, 344, 282

\bibitem[{{Marietta} {et~al.}(2000){Marietta}, {Burrows}, \&
  {Fryxell}}]{Marietta}
{Marietta}, E., {Burrows}, A., \& {Fryxell}, B. 2000, ApJS, 128, 615

\bibitem[{Nomoto {et~al.}(1984)Nomoto, Thielemann, \& Yokoi}]{Nomoto_w7}
Nomoto, K., Thielemann, F., \& Yokoi, K. 1984, ApJ, 286, 644

\bibitem[{Patat {et~al.}(1996)}]{Patat_94D}
Patat, F. {et~al.} 1996, MNRAS, 278, 111

\bibitem[{{Plewa} {et~al.}(2004){Plewa}, {Calder}, \& {Lamb}}]{Plewa_GCD}
{Plewa}, T., {Calder}, A.~C., \& {Lamb}, D.~Q. 2004, ApJ, 612, L37

\bibitem[{{Reinecke} {et~al.}(2002){Reinecke}, {Hillebrandt}, \&
  {Niemeyer}}]{Reinecke_3D}
{Reinecke}, M., {Hillebrandt}, W., \& {Niemeyer}, J.~C. 2002, A\&A, 391, 1167

\bibitem[{Thomas(2003)}]{Thomas_Thesis}
Thomas, R. 2003, PhD thesis, University of Oklahoma

\bibitem[{{Thomas} {et~al.}(2004)}]{Thomas_00cx}
{Thomas}, R.~C. {et~al.} 2004, ApJ, 601, 1019

\bibitem[{{Wang} {et~al.}(2003)}]{Wang_01el}
{Wang}, L. {et~al.} 2003, ApJ, 591, 1110

\end{thebibliography}

\end{document}